\documentclass[prd,onecolumn]{revtex4}
\usepackage{dcolumn}
\usepackage{multirow}
\usepackage{graphicx}
\usepackage{amssymb}
\usepackage{bm}
\usepackage{hyperref}
\usepackage{epstopdf}
\usepackage{color}
\usepackage{mathrsfs}
\usepackage{amsmath,amssymb,amsthm}
\usepackage{rotating}
\usepackage{sverb, longtable}
\usepackage{subfigure}

\usepackage[graphicx]{realboxes}
\usepackage{adjustbox}
\begin{document}

\title{Assessing the potential of cluster edges as a standard ruler on constraining dark energy models }
\author{Deng Wang}
\email{cstar@nao.cas.cn}
\affiliation{National Astronomical Observatories, Chinese Academy of Sciences, Beijing, 100012, China}

\begin{abstract}
We assess comprehensively the potential of galaxy cluster edges as a standard ruler in measuring cosmological distances and probing exotic physics. Confronting five alternative cosmological models with  cluster edges from the near future Dark Energy Spectroscopic Instrument survey, we conclude that cluster edges can serve as a promising probe to constrain models beyond $\Lambda$CDM. Especially,  
the constraining precision of equation of state of dark energy from cluster edges is just about two times larger than that from the Pantheon Type Ia supernovae sample. We find that the constraining power of cluster edges can be exhibited better by combining it with other probes. Combining cluster edges with cosmic microwave background, baryon acoustic oscillations, Type Ia supernovae, cosmic chronometers and simulated gravitational-wave events from the space-based Einstein Telescope, respectively, we constrain $\Lambda$CDM and find that the data combination of cosmic microwave background and cluster edges gives the best constraint on the Hubble constant $H_0$ with a $0.5\%$ precision and the matter density ratio $\Omega_m$ with a $1.6\%$ precision among these five pair datasets, and that the data combination of gravitational waves and cluster edges almost shares the same constraining power as that of Type Ia supernovae and cluster edges. We also give the most stringent constraint on $\Lambda$CDM by combining cluster edges with available cosmological data. With the help of other probes, galaxy cluster edges can give new insights on exotic physics better at cosmological scales.

\end{abstract}
\maketitle

\section{Introduction}
The standard cosmological model, $\Lambda$-cold dark matter ($\Lambda$CDM), has achieved great success in explaining the physical phenomenon ranging from large to small scales, since the late-time accelerated expansion of the universe is discovered by two independent Type Ia supernovae (SNe Ia) groups \cite{1,2}. However, this model is not as perfect as we imagine, and it faces at least two intractable problems, namely the coincidence and fine-tuning problems \cite{3}. The former is why energy densities of dark matter and dark energy are of the same order of magnitude at present, since their energy densities are so different from each other during the evolution of the universe, while the latter indicates the theoretical prediction for vacuum energy density is far larger than its observed value. Furthermore, during the recent several years, different cosmological experiments tell us that the $\Lambda$CDM model also faces at least two tensions, i.e., the Hubble constant ($H_0$) and matter fluctuation amplitude ($\sigma_8$) tensions. The $H_0$ tension is that indirectly derived value from the Planck-2018 cosmic microwave background (CMB) measurement \cite{4} under the assumption of $\Lambda$CDM is 4.4$\sigma$ lower than the direct measurement of present cosmic expansion rate from the Hubble Space Telescope (HST), while the $\sigma_8$ one indicates that present matter fluctuation amplitude in the linear regime measured by several low redshift probes including weak lensing \cite{5,6,7}, cluster counts \cite{8} and redshift space distortions \cite{9} is about 2$\sigma$ lower than that indirectly measured by the Planck-2018 CMB observation \cite{4}. All of this implies that $\Lambda$CDM may not be the underlying model governing the evolution of the universe at both the background and perturbation levels. Especially, the very severe $H_0$ tension as an important motivation inspires a great deal of works to propose a new cosmological model (dark energy or modified gravity) to replace $\Lambda$CDM \cite{10,11,12,13,14,15,16,17,18,19,20,21,22,23,24,25,26,a4,a5}. In general, the reason leading to this large background discrepancy can be ascribed to either possible systematic errors or new physics. It is noteworthy that, in practice, independent determinations on $H_0$ by other probes are very important complements to alleviate this tension or give some useful indications. For instance, the LIGO collaboration \cite{27} gives the first standard siren determination $H_0=70^{+12}_{-8}$ km s$^{-1}$ Mpc$^{-1}$ by using a combination of a distance estimation from gravitational wave observations and a Hubble velocity estimation from electromagnetic observations. It is a completely new information channel independent from the traditional electromagnetic ones, although the error of $H_0$ is very large at the current stage. 

Most recently, the authors in Ref.\cite{28} point out that galaxy cluster edges can act as a standard ruler to estimate $H_0$ and measure the cosmological distances. Using the feature that the spatial extent of a cluster is correlated with the amplitude of the velocity dispersion profile, one can infer the physical cluster size through the amplitude of the velocity dispersion data. As a consequence, these observations of angular scale can be transformed into distance data to clusters. Under the assumption that the relation between cluster velocity dispersion and cluster radius can be calibrated by simulations, they apply their method into the available Sloan Digital Sky Survey (SDSS) data and future Dark Energy Spectroscopic Instrument (DESI) data, and obtain $3\%$ and $1.3\%$ determinations on $H_0$, respectively. These estimations are very close to $1.9\%$ measurement ($H_0=74.04\pm1.42$ km s$^{-1}$ Mpc$^{-1}$) \cite{29} from HST and  $0.8\%$ measurement ($H_0=67.36\pm0.54$ km s$^{-1}$ Mpc$^{-1}$) \cite{4} from Planck-2018 CMB data, and can provide useful indications for the $H_0$ problem. 

Based on the concern that this newly proposed cosmological probe, cluster edges, gives a competitive estimation of $H_0$, we wonder that whether it can also give a good constraint on dark energy or modified gravity at the background level. Therefore, we place constraints on seven cosmological models including $\Lambda$CDM, $\omega$CDM ($\omega$ is the equation of state of dark energy), non-flat $\Lambda$CDM (o$\Lambda$CDM), interacting dark energy (IDE), dynamical dark energy (DDE) and $f(R)$ gravity by using the forecasted cluster edges data from the DESI survey. We find that cluster edges can act as a promising probe to constrain the cosmological models beyond $\Lambda$CDM, and that, among five pair combinations, the most precise estimations of $H_0$ and $\Omega_m$ are obtained by combining the Planck-2018 CMB observation with cluster edges data.            

This work is organized in the following manner. In the next section, we review the cosmological models used in this analysis. In Section III, we describe the galaxy cluster edges data and analysis methodology. In Section IV, we exhibit the numerical results. The discussions and conclusions are presented in the final section.

\section{Cosmological models}
In this section, we introduce seven cosmological models to be constrained by galaxy cluster edges data. Throughout this work, we investigate these seven models in a Friedmann-Robertson-Walker (FRW) universe in the framework of general relativity (GR), and just focus on the late-time cosmology, consequently neglecting the contribution from radiation in the cosmic pie. The homogeneous and isotropic universe described by the FRW metric
\begin{equation}
\mathrm{d}s^2=-dt^2+a^2(t)\left[\frac{\mathrm{d}r^2}{1-Kr^2}+r^2\mathrm{d}\theta^2+r^2sin^2\theta \mathrm{d}\phi^2\right],      \label{1}
\end{equation}  
where $a(t)$ and $K$ are the scale factor at cosmic time $t$ and the Gaussian curvature of spacetime, respectively. Substituting Eq.(\ref{1}) into the gravitational field equation, one can have the so-called Friedmann equations as follows
\begin{equation}
H^2=\frac{8\pi G}{3}\Sigma\rho_i,     \label{2}
\end{equation}   
\begin{equation}
\frac{\ddot{a}}{a}=-\frac{4\pi G}{3}\Sigma(\rho_i+3p_i),     \label{3}
\end{equation}   
where $H$ is the Hubble parameter, and $\rho_i$ and $p_i$ represent the mean energy density and mean pressure of different components. Specifically, in this analysis, we consider the matter and dark energy components in the cosmic pie. Combining Eqs.(\ref{2}) with (\ref{3}), one can easily obtain the dimensionless Hubble parameter (DHP), which characterizes the background evolution of a cosmological model, for the $\Lambda$CDM model
\begin{equation}
E_{\mathrm{\Lambda CDM}}(z)=\left[\Omega_{m}(1+z)^3+1-\Omega_{m}\right]^{\frac{1}{2}}, \label{4}
\end{equation}
while for the o$\Lambda$CDM model considering the non-zero curvature of the universe, it is expressed as
\begin{equation}
E_{\mathrm{o\Lambda CDM}}(z)=\left[\Omega_{m}(1+z)^3+\Omega_{K}(1+z)^2+1-\Omega_{m}-\Omega_{K}\right]^{\frac{1}{2}}, \label{5}
\end{equation}
where $z$, $\Omega_m$ and $\Omega_K$ denote the redshift, present matter density ratio and cosmic curvature, respectively. Note that here the equation of state of dark energy has been fixed to -1 \cite{4}. 

To explore the property of dark energy, the simplest model is $\omega$CDM, a one-parameter extension to $\Lambda$CDM, where dark energy is described by a barotropic fluid with a constant equation of state $\omega(z)=\omega$. The DHP for the spatially flat $\omega$CDM model is shown as
\begin{equation}
E_{\mathrm{\omega CDM}}(z)=\left[\Omega_{m}(1+z)^3+(1-\Omega_{m})(1+z)^{3(1+\omega)}\right]^{\frac{1}{2}}. \label{6}
\end{equation}
It is obvious that $\omega$CDM with $\omega=-1$ reduces to $\Lambda$CDM. 

Furthermore, it is natural to study the one-parameter extension to $\omega$CDM, i.e., the so-called CPL Chevallier-Polarski-Linder) parameterization \cite{a1,a2}. Its DHP is written as

\begin{equation}
E_{\mathrm{CPL}}(z)=\left[\Omega_{m}(1+z)^3+(1-\Omega_{m})(1+z)^{3(1+\omega_0+\omega_a)\mathrm{e}^{\frac{-3\omega_az}{1+z}}}\right]^{\frac{1}{2}}, \label{e1}
\end{equation}

where $\omega_0=-1$ and $\omega_a=0$ reduce to $\Lambda$CDM.

Since the cosmic acceleration is discovered, one of the most important topics in the field of modern cosmology is whether the dark energy evolves over time or not. To address this problem, we have proposed a DDE model based on phenomenological dark energy density parametrization in our previous work \cite{14}, and its corresponding DHP can be expressed as 
\begin{equation}
E_{\mathrm{DDE}}(z)=\left[\Omega_{m}(1+z)^{3}+(1-\Omega_{m})(1+\beta-\frac{\beta}{1+z})\right]^{\frac{1}{2}},   \label{7}
\end{equation}
where $\beta$ denotes the unique parameter characterizing this parametrization model. It is easy to find that this DDE model reduces to $\Lambda$CDM when $\beta=0$, and that if $\beta$ has any departure from zero, dark energy will be dynamical. 

Because the nature of dark energy and dark matter is still unclear, to solve the cosmological puzzles, an important conjecture that dark energy can interact with dark matter emerges. It is interesting to study whether there is an interaction between dark energy and dark matter in the dark sector by using the galaxy cluster edges observations. 

Taking a modified matter expansion rate $\epsilon$ into account, the DHP of IDE model \cite{30} to be confronted with data is written as
\begin{equation}
E_{\mathrm{IDE}}(z)=\left[\frac{3\Omega_{m}}{3-\epsilon}(1+z)^{3-\epsilon}+1-\frac{3\Omega_{m}}{3-\epsilon}\right]^{\frac{1}{2}},   \label{8}
\end{equation}
where the typical parameter $\epsilon<0$ indicates that the momentum transfers from dark matter to dark energy and vice versa.

We are also interested in investigating the ability of cluster edges data in constraining the modified gravity model, which normally uses a curvature fluid to explain the cosmic acceleration. Specifically, we will consider the $f(R)$ gravity, where the Ricci scalar $R$ in the Lagrangian of GR is generalized to a function of $R$. The Hu-Sawicki $f(R)$ gravity (hereafter HS) model \cite{31} reads as    
\begin{equation}
f(R)=R-\frac{m^2 c_1(\frac{R}{m^2})^n}{1+c_2(\frac{R}{m^2})^n},   \label{9} 
\end{equation}  
which can be rewritten as 
\begin{equation}
f(R)=R-\frac{2\Lambda}{1+(\frac{b\Lambda}{R})^n},  \label{10}
\end{equation} 
where $b=\frac{2c_2^{1-1/n}}{c_1}$ and  $\Lambda=\frac{m^2c_1}{2c_2}$. It is obvious that three parameters $c_1$, $c_2$ and $n$ are transformed into two ones $b$ and $n$ in this model.
For simplicity, we choose $n=1$. Hence, one can easily find that the HS model reduces to $\Lambda$CDM when $b=0$. If the value of $b$ has a deviation from zero, GR needs to be modified.  
Following the method which uses Taylor series expansion to express the Hubble parameter $H_{HS}(z)$ around $b=0$ in Ref.\cite{32}, we take this good approximation to implement constraints on the HS $f(R)$ model by using the cluster edges data. The specific formula can be found in the appendix of Ref.\cite{32}.

Note that, for the purpose of exploring the property of cosmic acceleration, the cosmological models we consider here can act as a good representative of non-standard cosmologies.

\section{Cosmological data and methodology}
In Ref.\cite{28}, through studying the radial dependence of the stacked line-of-sight velocity dispersion of galaxies relative to the central galaxy of SDSS redMaPPer clusters, the authors find a kink feature, which can serve as a standard ruler. The kink is interpreted as spatial extent of galaxies orbiting the redMaPPer clusters, and this transition scale can act as the galaxy cluster edge radius. Their key idea about using cluster edges as a standard ruler is that more massive halos occupy more space, i.e., halos with larger line-of-sight velocity dispersions have larger radii. Their analysis is based on an assumption that numerical simulations can help calibrate a scaling relation between the velocity dispersion $\sigma_{v}$   and edge radius $R_{\mathrm{edge}}$ of a halo
\begin{equation}
R_{\mathrm{edge}}=A\sigma_{v}^\alpha, \label{11}
\end{equation}
where $A$ and $\alpha$ are two parameters to be calibrated from simulations. Using the spectroscopic observations, one can empirically measure $\sigma_{v}$ and, consequently, determine $R_{\mathrm{edge}}$. Because the spectroscopic observations can also measure the velocity dispersion profile as a function of angular separation $\theta$, the transition scale $\theta_{kink}$ corresponding the kink feature can be determined. Then, the angular diameter distance can be written as 
\begin{equation}
D_A(z_{\mathrm{cluster}})=\frac{R_\mathrm{edge}}{\theta_\mathrm{kink}}, \label{12}
\end{equation} 
where $z_{\mathrm{cluster}}$ denotes the redshift of a observed galaxy cluster.
This is the key formula we use to derive constraints on cosmological models beyond $\Lambda$CDM. 

It is worth noting that the authors in Ref.\cite{28} use a full forward model to overcome several difficulties encountered in their method. Overall, there are several possible elements to affect the final estimation on $R_{\mathrm{edge}}$: (i) in order to estimate the constraining power of galaxy sample, an assumption that linking the orbiting velocity dispersion to the edge radius are perfectly calibrated is made; (ii) a remarkable effect, cluster selection, will lead to a large uncertainty on the final error of recovered $H_0$; (iii) calibrating extra halo structural parameters, the galaxy density profiles and the in-fall velocity dispersion ones, will possibly produce significant improvements to final forecast. The detailed information can be found in their work \cite{28}. 

\begin{figure}[htbp] 
	\centering
	\includegraphics[scale=0.7]{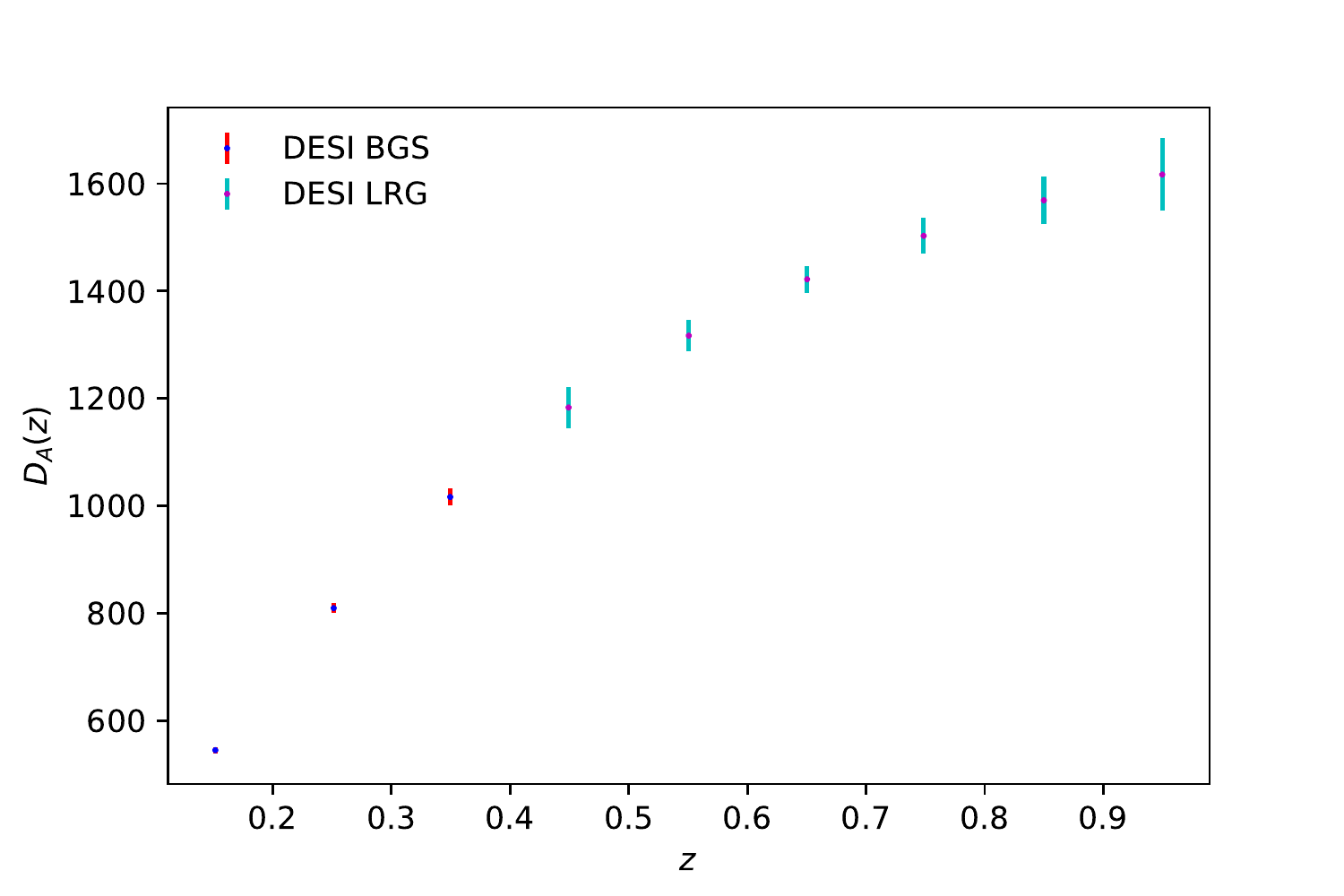}
	\caption{ The forecasted angular diameter distance data of the DESI survey is shown, consisting of the BGS sample in the redshift range $z\in[0.1,0.4)$ and the LRG one at redshifts $z\geqslant0.4$.  }
	\label{f1}
\end{figure}

Applying the above method into the near future DESI survey, they obtain the galaxy cluster edges data, which is shown in Fig.\ref{f1}. This dataset consisting of the DESI Bright Galaxy Sample (BGS) in redshifts $z\in[0.1,0.4)$ and the DESI Luminous Red Galaxy (LRG) sample at redshifts $z\geqslant0.4$, is actually the angular diameter distance data characterizing the background evolution of the universe. Notice that this mocked dataset is created by assuming the fiducial $\Lambda$CDM cosmology $\Omega_m=0.3$ and $H_0=70$ km s$^{-1}$ Mpc$^{-1}$. 
This dataset is denoted as ``E'' hereafter.

Besides the new cluster edges data, we will also consider the other five cosmological probes encompassing the CMB, baryon acoustic oscillations (BAO), SNe Ia, cosmic chronometers (CC) and gravitational waves (GW) in this analysis, which are introduced in the following context. 

{\it CMB}: The Planck CMB experiment plays an extremely important role in measuring many aspects of formation and evolution of the universe such as matter components, topology and large scale structure effects. In principle, here we shall employ the original CMB temperature and polarization data to constrain a cosmological model. However, since we just focus on studying the evolution of the universe at the background level, we use the distance-related information instead. To save computational effort, in this work, we use the distance prior from Planck-2018 TTTEEE$+$lowE$+$lensing data, i.e., compressed CMB data from Ref.\cite{4} to implement the cosmological constraints. Hereafter we refer to this data as ``C''. 

{\it BAO}: BAO is a clean probe to study the evolution of the universe, which is unaffected by uncertainties in the nonlinear evolution of the matter density field and other systematic uncertainties which may affect other observations. Measuring the position of these oscillations in the matter power spectrum at different redshifts can constrain the background evolution of the universe after decoupling and break degeneracies between parameters better. We adopt the 6dFGS sample at the effective redshift $z_{eff}=$ 0.106 \cite{33}, the SDSS-MGS one at $z_{eff}=$ 0.15 \cite{34} as well as the BOSS DR12 data at three effective redshifts $z_{eff}=$ 0.38, 0.51 and 0.61 \cite{35}. This dataset is identified as ``B''. 

{\it SNe Ia}: SNe Ia, the so-called standard candle, is a powerful distance indicator to probe the background dynamics of the universe, particularly, the Hubble parameter and equation of state of dark energy. In this analysis, we use the largest SNe Ia ``Pantheon'' sample today, which integrates the SNe Ia data from the Pan-STARRS1, SNLS, SDSS, low-z and HST surveys and contains 1048 spectroscopically confirmed data points in redshifts $z \in [0.01, 2.3]$ \cite{36}. We denote this dataset as ``S''.

{\it} CC: To study the late-time evolution of the universe, we also include the CC as a complementary probe, which is independent of a cosmological model. Specifically, we employ 31 chronometers to carry out constraints on the background cosmology \cite{37}. We call this dataset as ``H''.    

{\it GW}: GW is the ripple in the fabric of spacetime generated by the acceleration of astrophysical objects. 
The GW standard sirens can also serve as a promising probe to help explore the nature of dark energy or modified gravity. Since standard sirens are self-calibrating, the luminosity distance $d_L$ of a source can be inferred directly from the observed GW signal, without the help of a cosmic distance ladder. It is interesting to compare the constraining power of galaxy cluster edges data with that of GW. Specifically, we consider the binary mergers of a neutron star with either a neutron star (BNS) or a black hole (NSBH) in the third-generation ground-based detector, the Einstein Telescope (ET), which probes the high-frequency GW events. In our simulation, the errors of luminosity distances can be divided into two parts: the instrumental error $\sigma_i$ which is determined by the Fisher matrix of luminosity distances, and an error $\sigma_l$ due to effects of weak lensing. Therefore, the final error on $d_L$ is $\sigma_{d_L}=\sqrt{\sigma_i^2+\sigma_l^2}$. The expected events rate of BNS and NSBH for ET per year is of order $10^3-10^7$. Nonetheless, only a small fraction ($\sim10^3$) can be observed in order to satisfy the constraint that GW events can be accompanied with the observation of a short gamma-ray burst due to the narrow beaming angle. This means that, if assuming the events rate is $10^5$ per year, we will obtain $10^2$ GW events with short gamma-ray bursts. Here we adopt the relatively optimistic case, i.e., following the method used in Ref.\cite{38}, we simulate 1000 GW events from the Einstein Telescope in the redshift range $z\in[0,5]$ under the assumption of $\Lambda$CDM. We refer the readers to the detailed simulation procedures in Ref.\cite{38}. 

To implement the standard Beyesian statistics and infer the posterior probability distribution of cosmological parameters, we use the publicly online package {\it EMCEE} \cite{a3}, which is an extensible pure-python Affine Invariant Markov chain Monte Carlo (MCMC) Ensemble sampler. To analyze the MCMC chains, we employ the available package {\it GetDist}.    

To assess the constraining power of cluster edges data, first of all, we place constraints on seven cosmological models using cluster edges alone. Subsequently, we constrain $\Lambda$CDM by combining cluster edges with CMB, BAO, SNe Ia, CC, and GW, respectively. Finally, we implement a joint constraint on $\Lambda$CDM using a data combination of CBSHE.

\section{Numerical results}
\begin{figure}[htbp]
	\centering
	\includegraphics[scale=0.76]{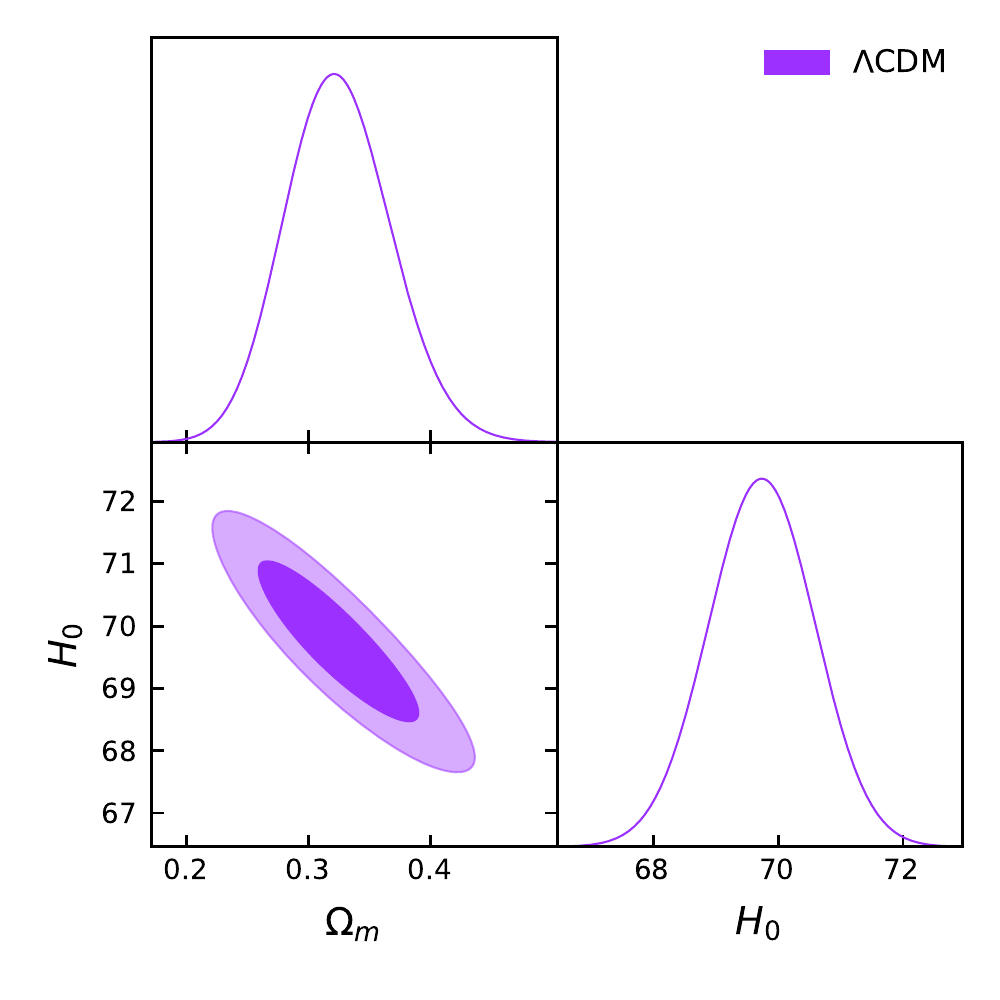}
	\includegraphics[scale=0.5]{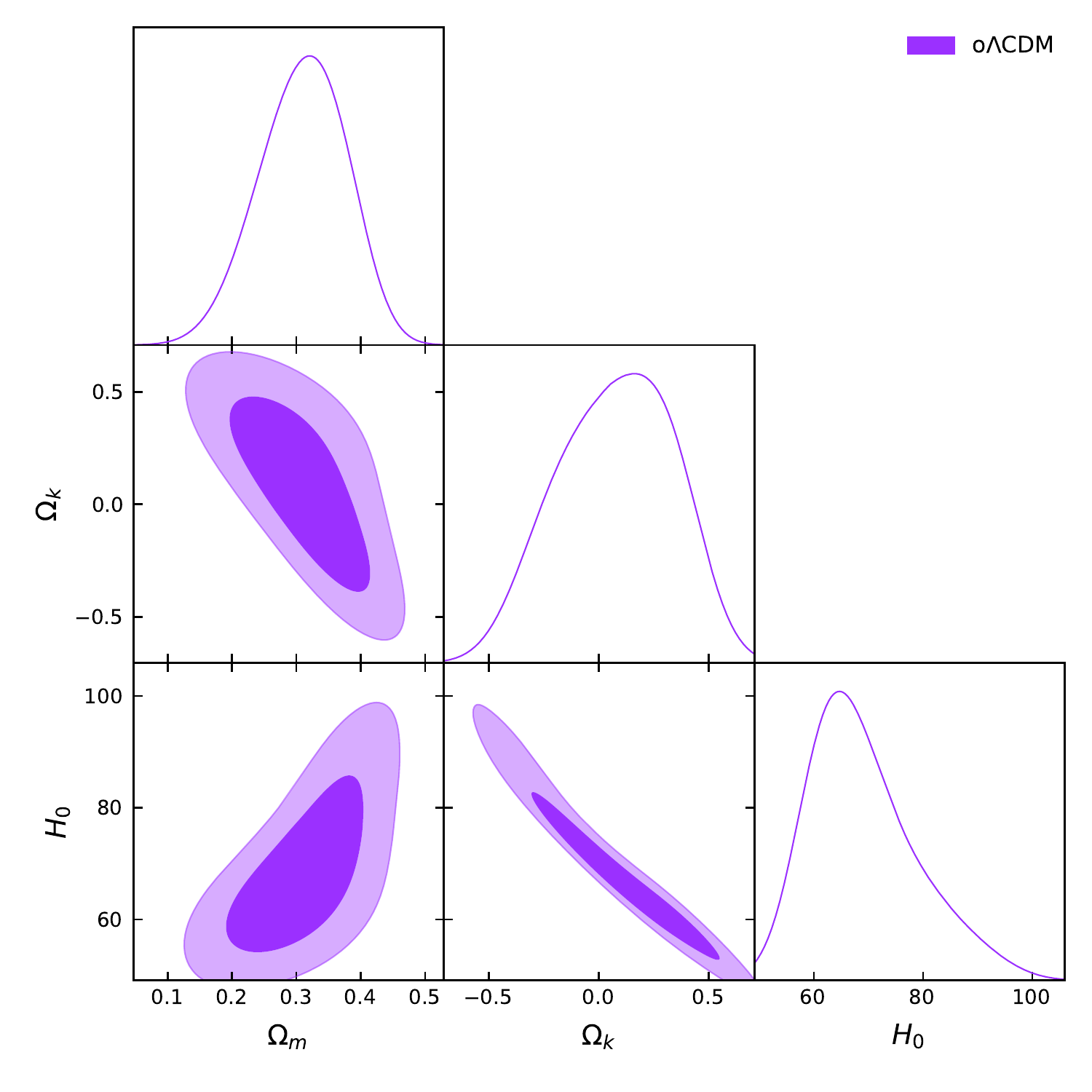}
	\includegraphics[scale=0.5]{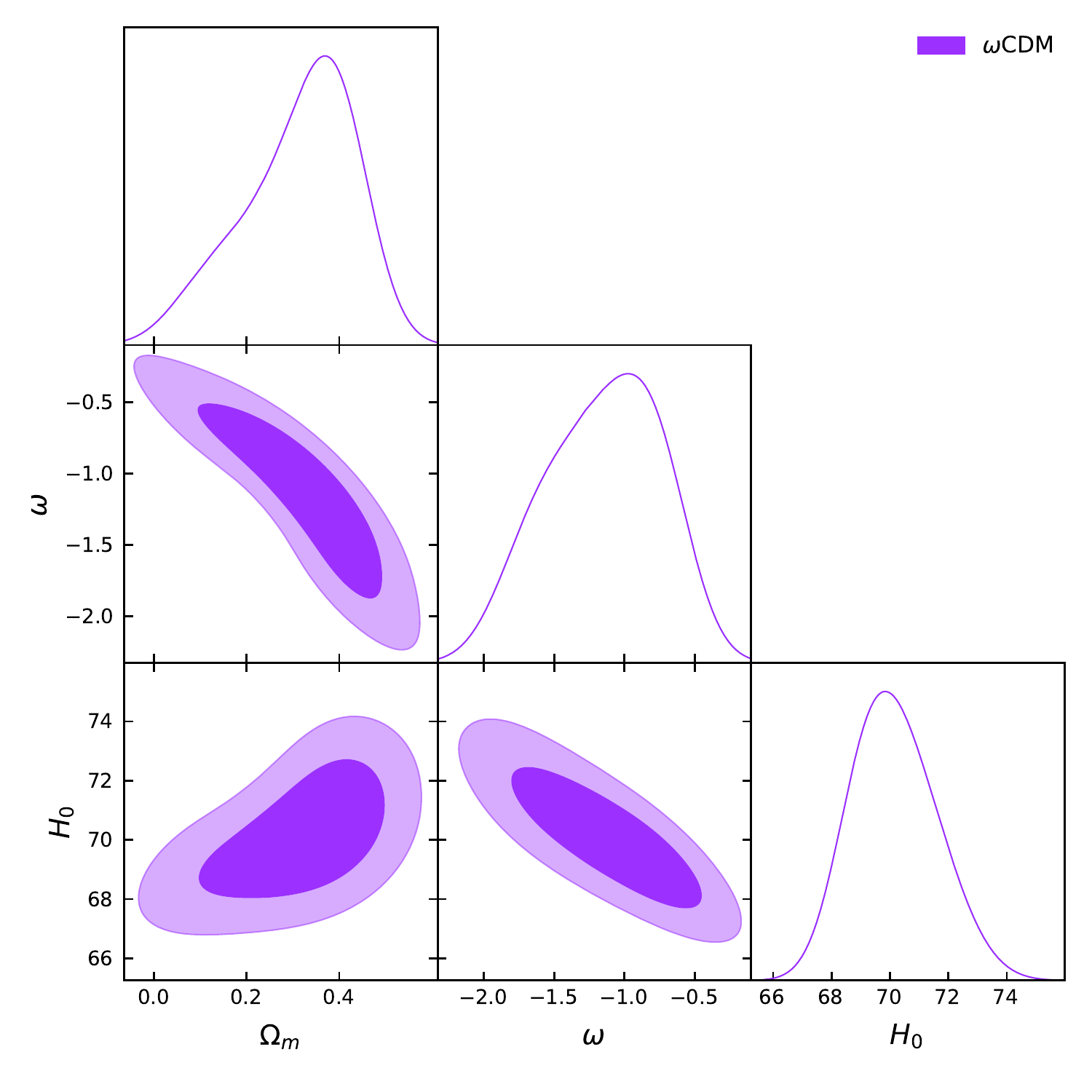}
	\includegraphics[scale=0.5]{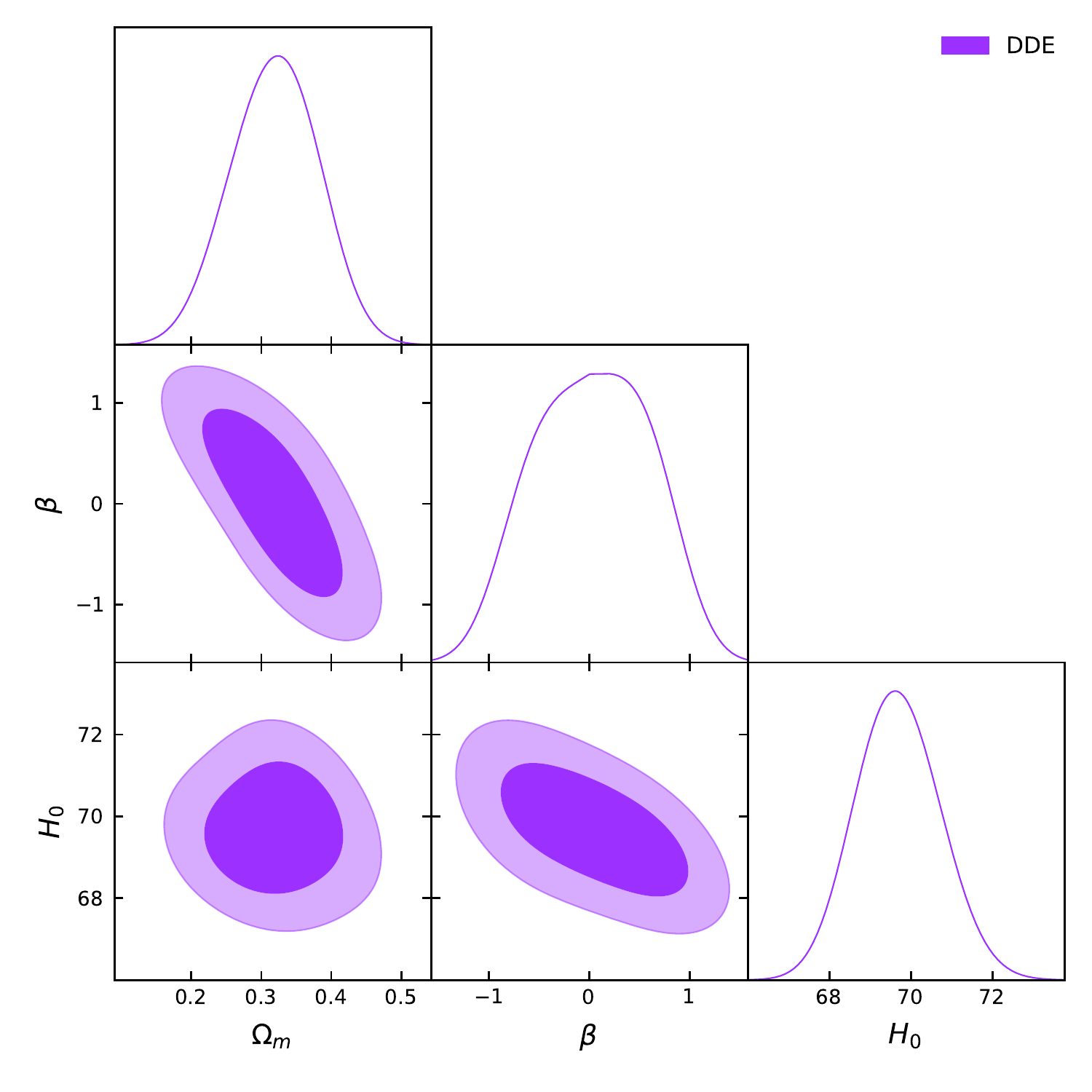}
	\includegraphics[scale=0.5]{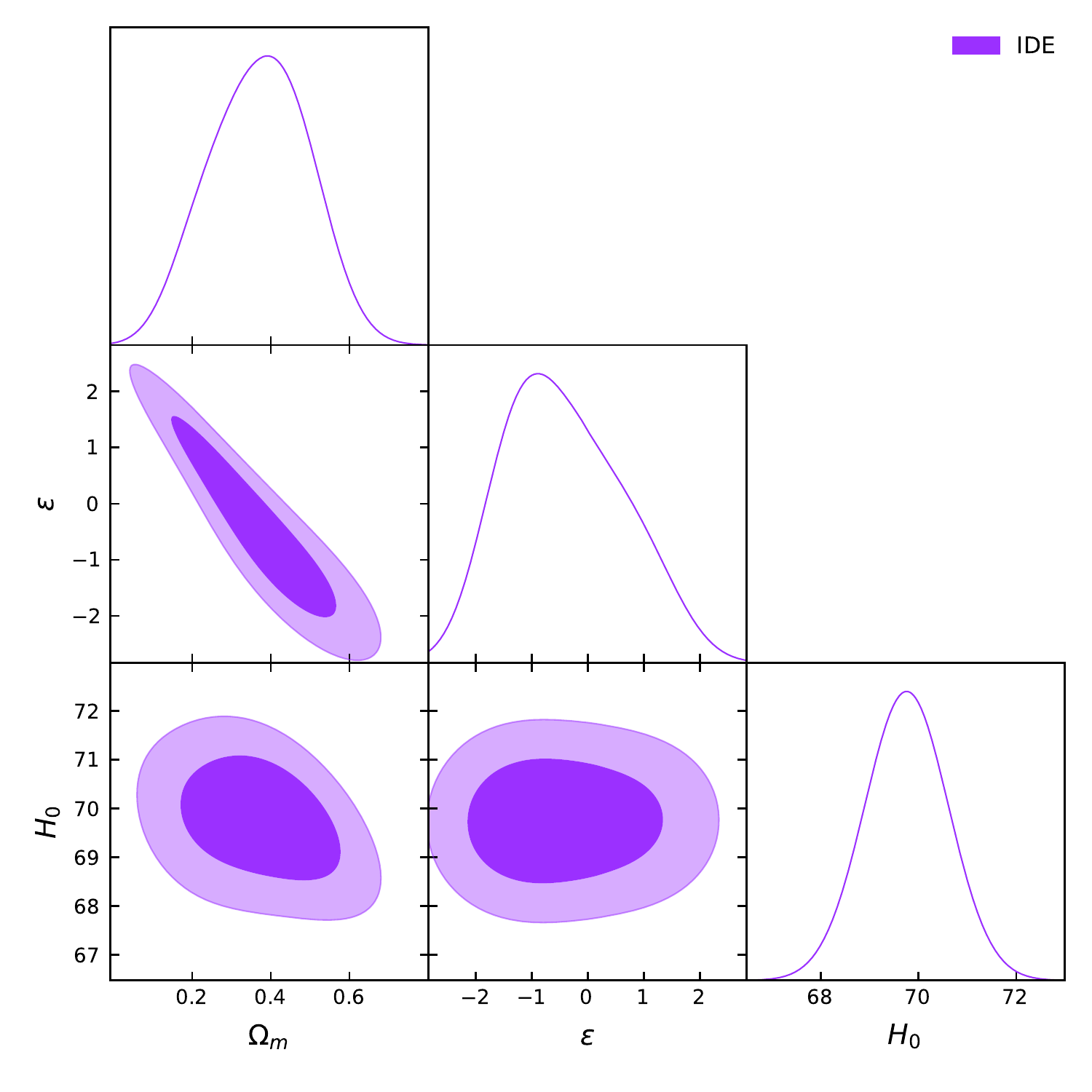}
	\includegraphics[scale=0.5]{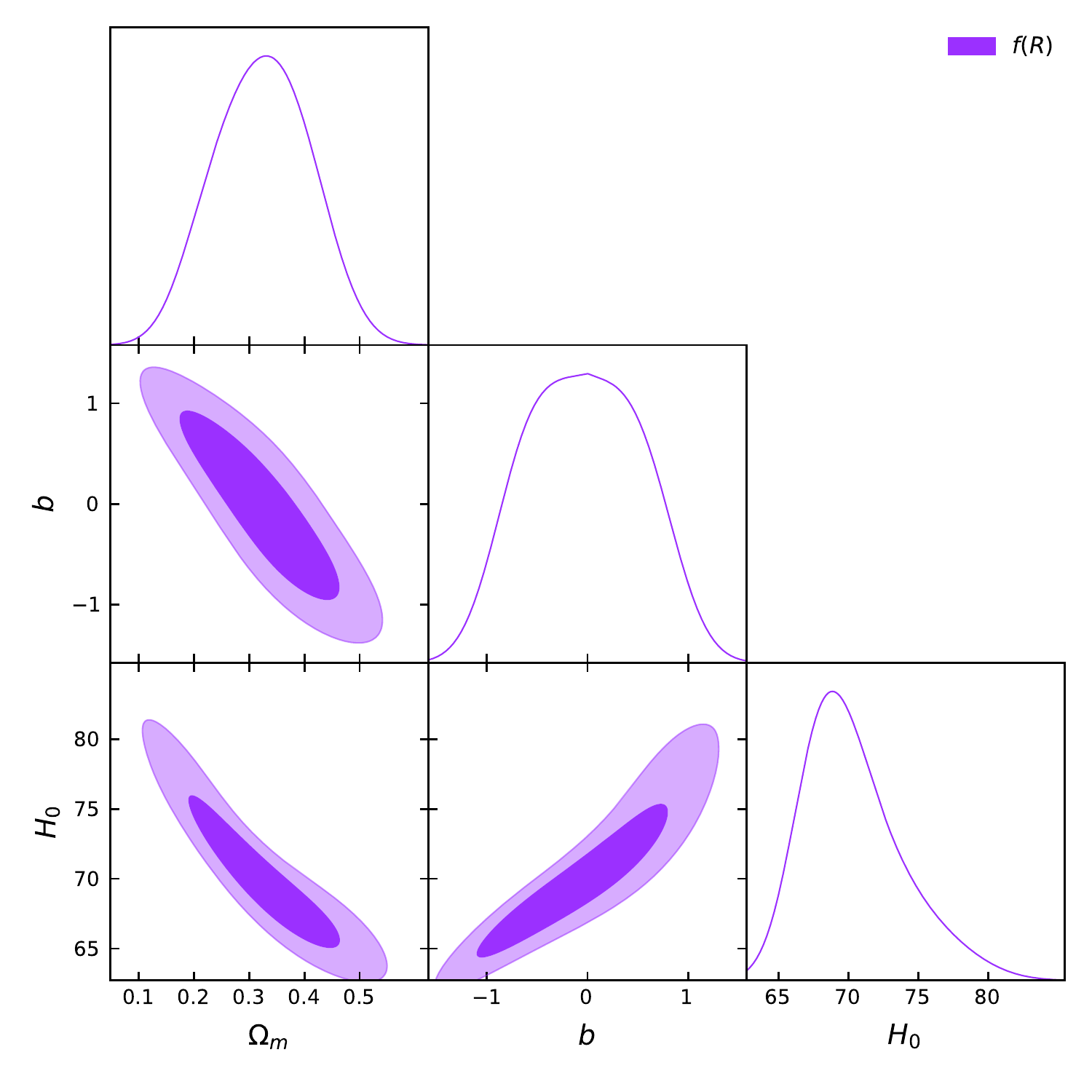}
	\caption{The marginalized constraints on the $\Lambda$CDM, o$\Lambda$CDM, $\omega$CDM, DDE, IDE and HS $f(R)$ models are shown by only using the galaxy cluster edges data from the near future DESI survey.  }
	\label{f2}
\end{figure}

Since cluster edges data can give a relatively tight constraint on $H_0$, we are full of interests to explore whether this probe can give new insights on new physics at cosmological scales. First of all, we constrain seven cosmological models with cluster edges only and the corresponding results are shown in Fig.\ref{f2} and Tab.\ref{t1}.

\begin{figure}[htbp]
	\centering
	\includegraphics[scale=0.5]{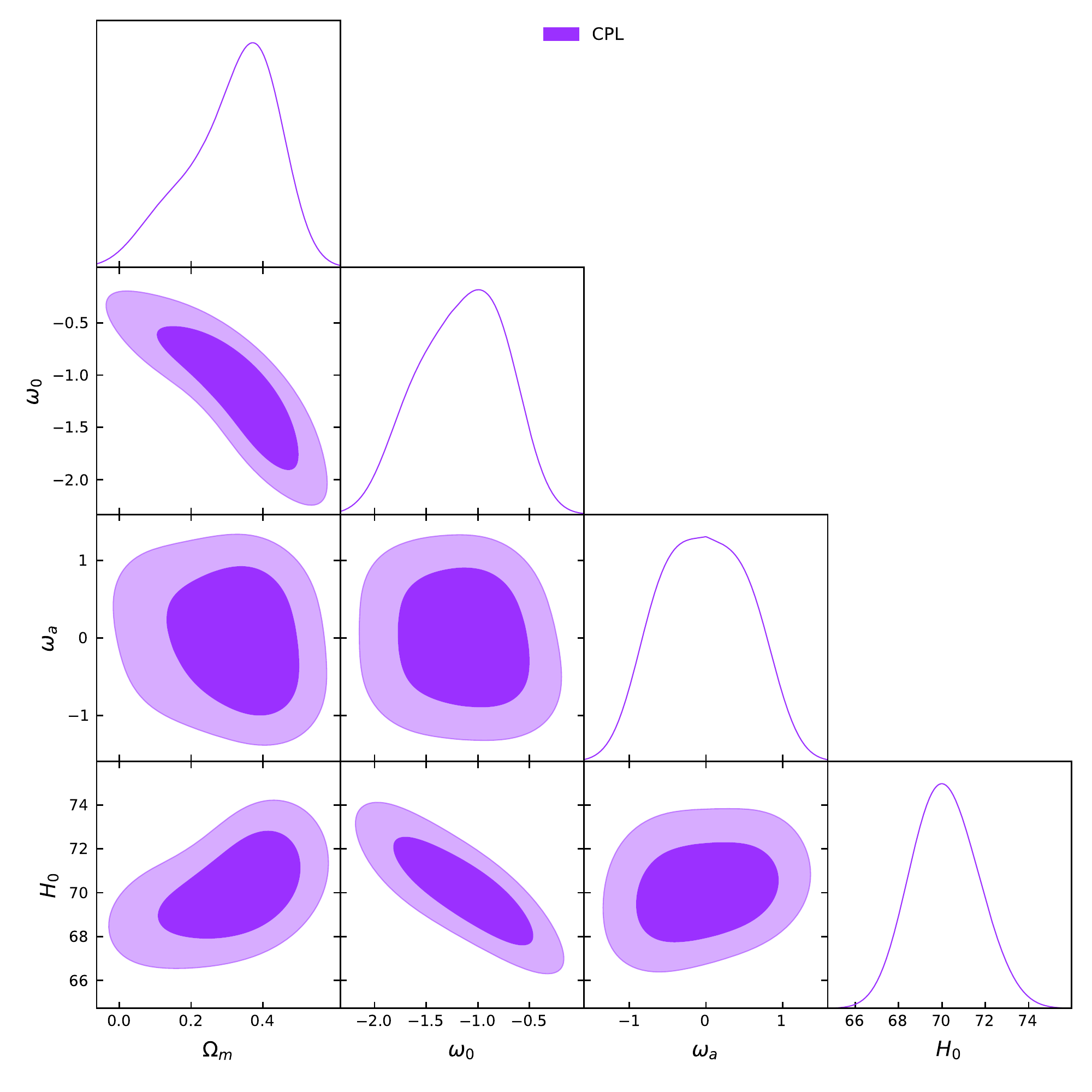}
	\caption{The marginalized constraint on the CPL model are shown by only using the galaxy cluster edges data from the near future DESI survey.  }
	\label{fadd}
\end{figure}

\begin{table}[h]
	\renewcommand\arraystretch{1.5}
	\caption{The constraining results of free parameters of seven cosmological models from galaxy cluster edges data alone are shown. }
	{\begin{tabular}{l |c| c |c| c| c | c |c } \toprule
			Parameters            &$\Lambda$CDM           & o$\Lambda$CDM       &$\omega$CDM                   &DDE           &IDE    &$f(R)$   & CPL               \\ \colrule
			$H_0$              &69.73$\pm$0.83   &$68.7^{+7.2}_{-12.0}$      &$70.1^{+1.4}_{-1.6}$        &  69.7$\pm$1.0  &69.75$\pm$0.84        &$70.2^{+2.6}_{-4.2}$    &  70.1$\pm$1.5       \\  
			$\Omega_{m}$       &$0.325\pm0.042$  &$0.311^{+0.073}_{-0.064}$  &$0.316^{+0.150}_{-0.093}$   &$0.319\pm0.062$ &$0.37\pm0.13$         &0.324$\pm$0.086      &  $0.318^{+0.15}_{-0.10}$            \\
			$\Omega_{K}$       &---              &$0.08^{+0.30}_{-0.25}$     &---                         &---             &---                   &---               &---               \\
			$\omega$           &---              &---                        &$-1.14^{+0.49}_{-0.39}$     &---             &---                   &---                &---             \\
			$\beta$            &---                        &---                      &---          &---         &0.03$\pm$0.57   &---                   &---                             \\
			$\epsilon$         &---                        &---                      &---                 &---             &$-0.43^{+0.95}_{-1.30}$    &---             &---              \\
			$b$                &---                        &---                      &---                 &---             &---                        &-0.04$\pm$0.57     &---              \\
			$\omega_0$         &---                        &---                      &---                 &---             &---                   &---             & $-1.15^{+0.48}_{-0.40}$\\
			$\omega_a$         &---                        &---                      &---                 &---             &---                   &---             & $-0.02\pm0.57$   \\
			\botrule
		\end{tabular}
		\label{t1}}
\end{table}

In $\Lambda$CDM, we can recover the prediction of $H_0$ with about $1.3\%$ precision in Ref.\cite{28}, which demonstrates that our method is valid. For the non-flat $\Lambda$CDM model, we find that the Hubble expansion rate $H_0$ can not be well constrained and suffers a very large error. Similarly, due to observational precision limitations and statistically small samples, the cosmic curvature $\Omega_K=0.08^{+0.30}_{-0.25}$ has a large error, although it is still consistent with zero. For the case of $\omega$CDM, we obtain a $H_0$ error $\sigma_{H_0}=1.6$ km s$^{-1}$ Mpc$^{-1}$, which is two times larger than that in $\Lambda$CDM. Meanwhile, we acquire the constraint on the equation of state of dark energy $\omega=-1.14^{+0.49}_{-0.39}$, a $43\%$ prediction, which is about two times larger than the $20\%$ precision from the Pantheon SNe Ia sample. Hence, cluster edges can be a complementary probe over the traditional standard candle SNe Ia.
For the CPL parameterization, we obtain a very similar constraint on $H_0$, $\Omega_m$ and $\omega_0$ compared to the case of $\omega$CDM.
One can find that the constraint on the key parameter $\omega_a=-0.02\pm0.57$ gives a very large uncertainty on the dynamical property of dark energy (see Fig.\ref{fadd}).  
For the DDE model, we obtain relatively stable values of $H_0$ and $\Omega_{m}$ compared to $\Lambda$CDM, and dynamical parameter $\beta=0.03\pm0.57$, which implies that there is no dynamical dark energy at the $1\sigma$ confidence level. For the IDE case, wo find a almost same $H_0$ value as $\Lambda$CDM, and a loose constraint on the interaction parameter $\epsilon=-0.43^{+0.95}_{-1.30}$, the error of which is far larger than our previous result $\epsilon=-0.00029^{+0.00028}_{-0.00025}$ \cite{39} from a joint constraint of CBSH and CMB lensing data. At the same time, due to the large error, this result also indicates that future DESI cluster edges data can not determine whether an interaction between dark matter and dark energy exist in the dark sector. For the HS $f(R)$ gravity model, we obtain a $6\%$ determination on $H_0$, which is about 5 times larger than the accuracy in $\Lambda$CDM. Similarly, the DESI data implies can not give a signal of modified gravity at the background level based on the larger error of the sole parameter $b$. In total, there is no new physics found in the near future galaxy cluster edges data from the DESI survey. Besides the $H_0$ values, interestingly, most $\Omega_m$ values are stable, but IDE meets unstable $\Omega_m$ one $0.37\pm0.13$, a $35\%$ determination. This can be ascribed to poor constraint on the interaction $\epsilon$, which has a strong correlation with the matter density $\Omega_m$.   

\begin{figure}[h]
	\centering
	\includegraphics[scale=0.55]{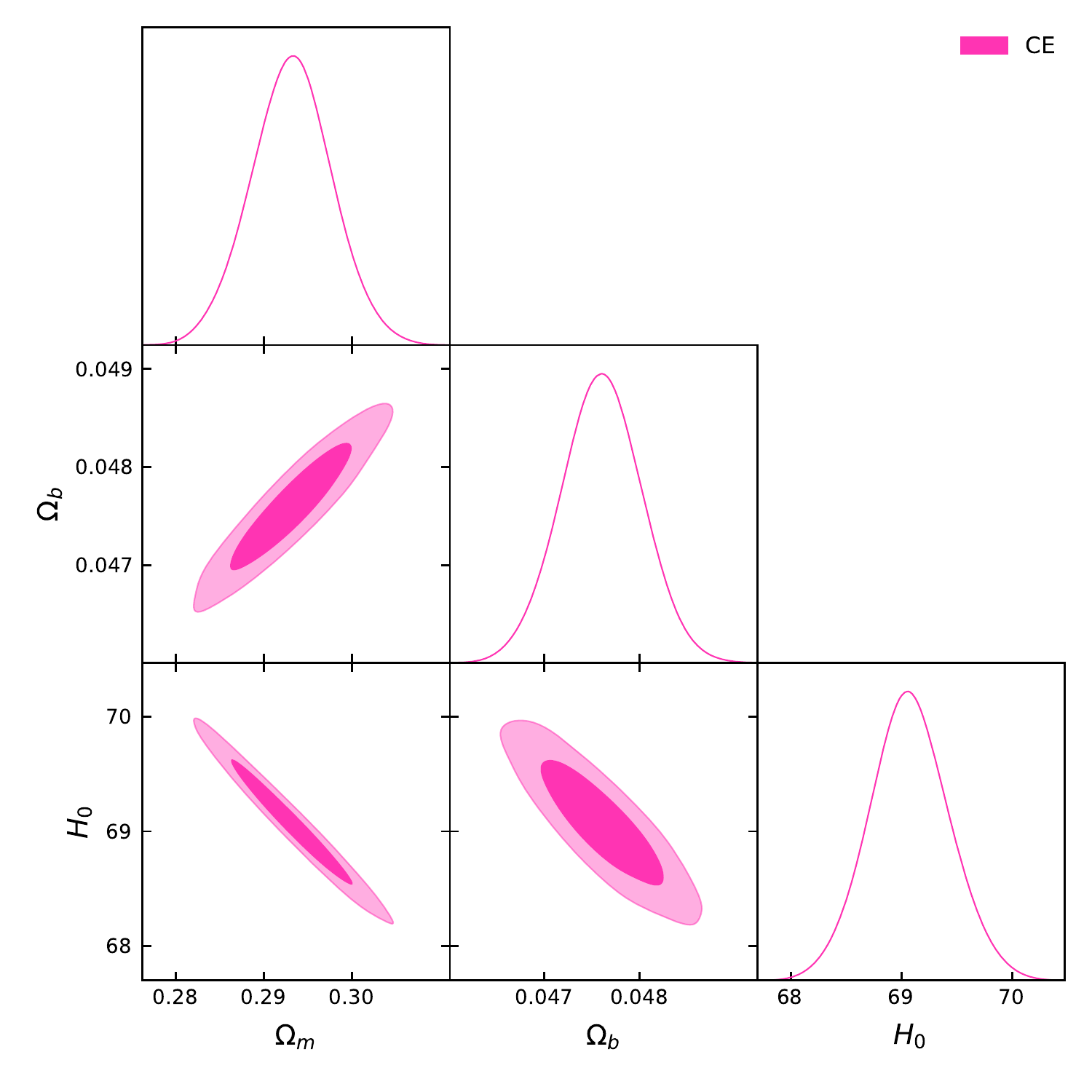}
	\includegraphics[scale=0.55]{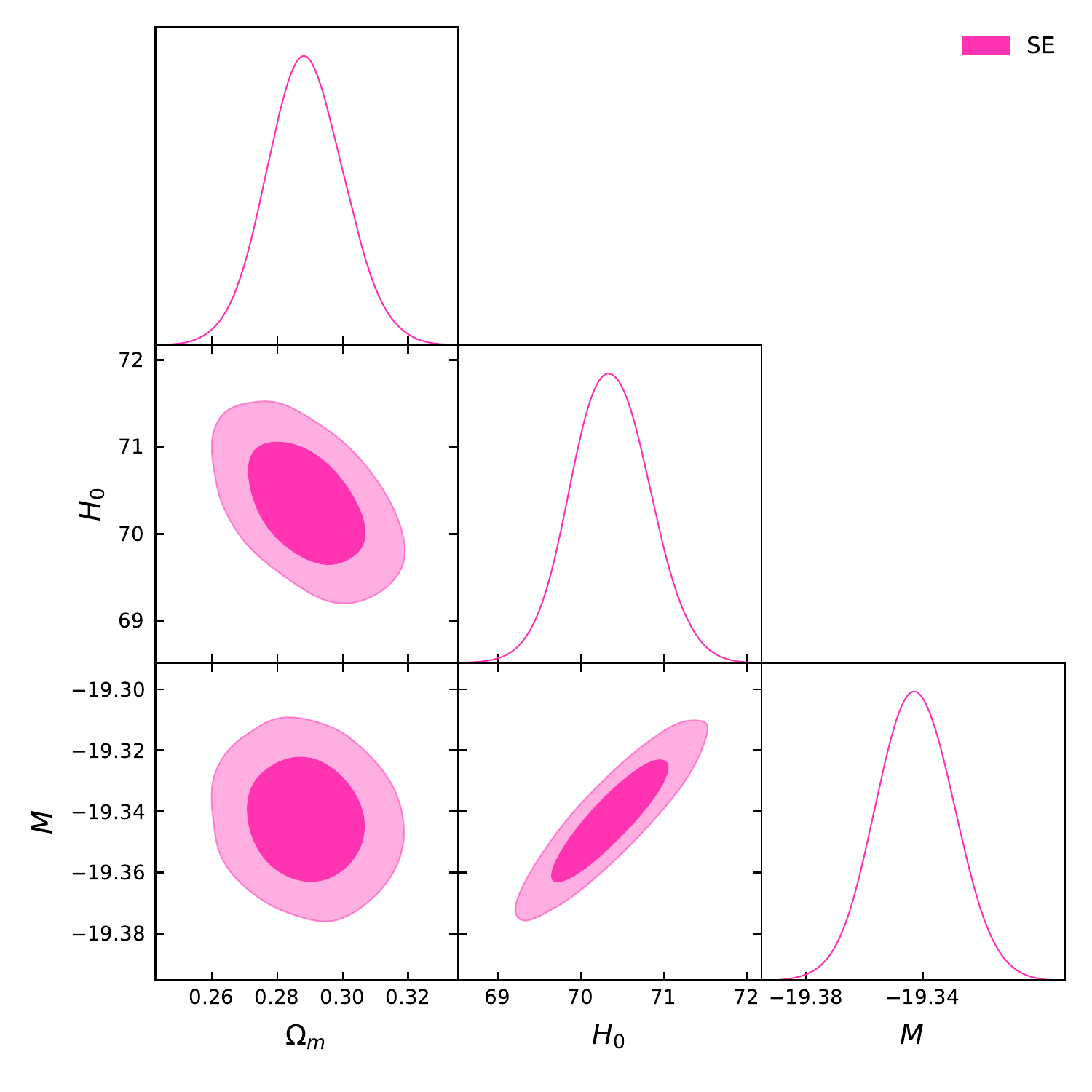}
	\caption{ The marginalized constraints on the $\Lambda$CDM model are shown by combining galaxy cluster edges data with CMB (left) and SNe Ia (right), respectively. }
	\label{f3}
\end{figure}

\begin{figure}[h]
	\centering
	\includegraphics[scale=0.55]{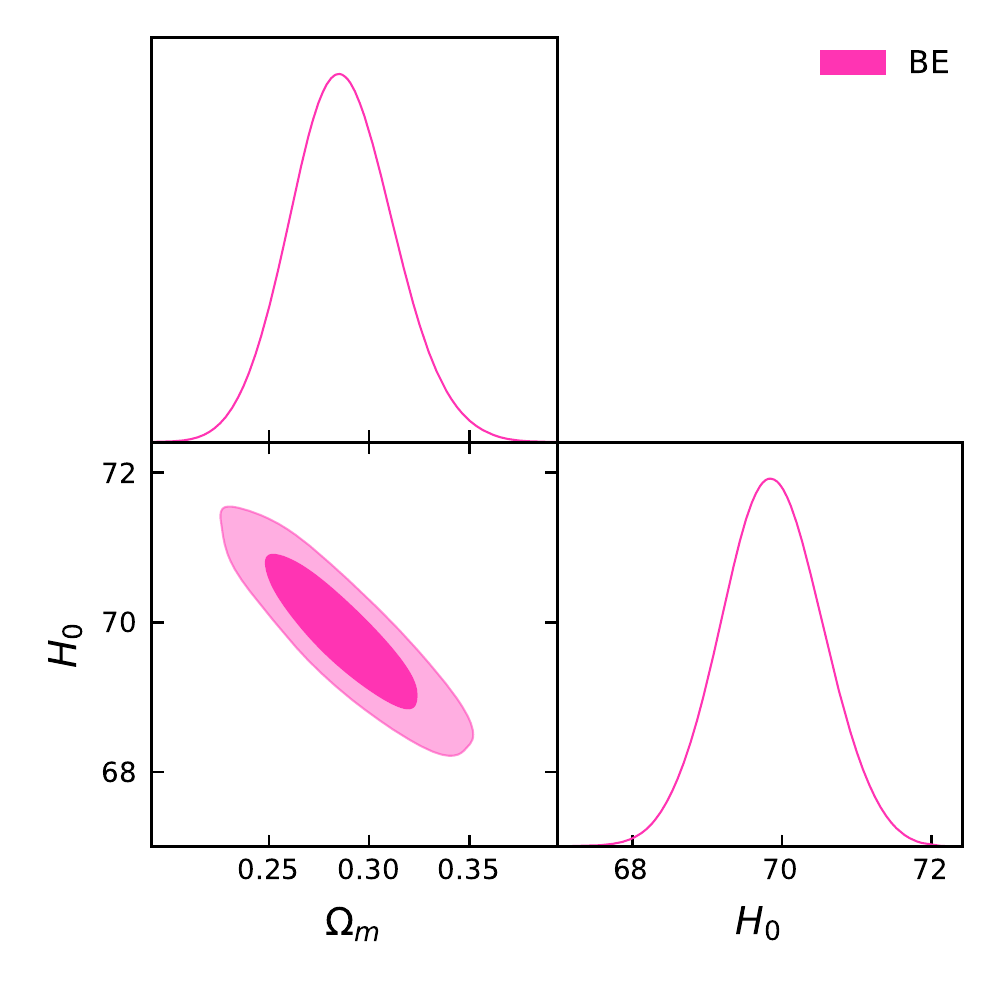}
	\includegraphics[scale=0.55]{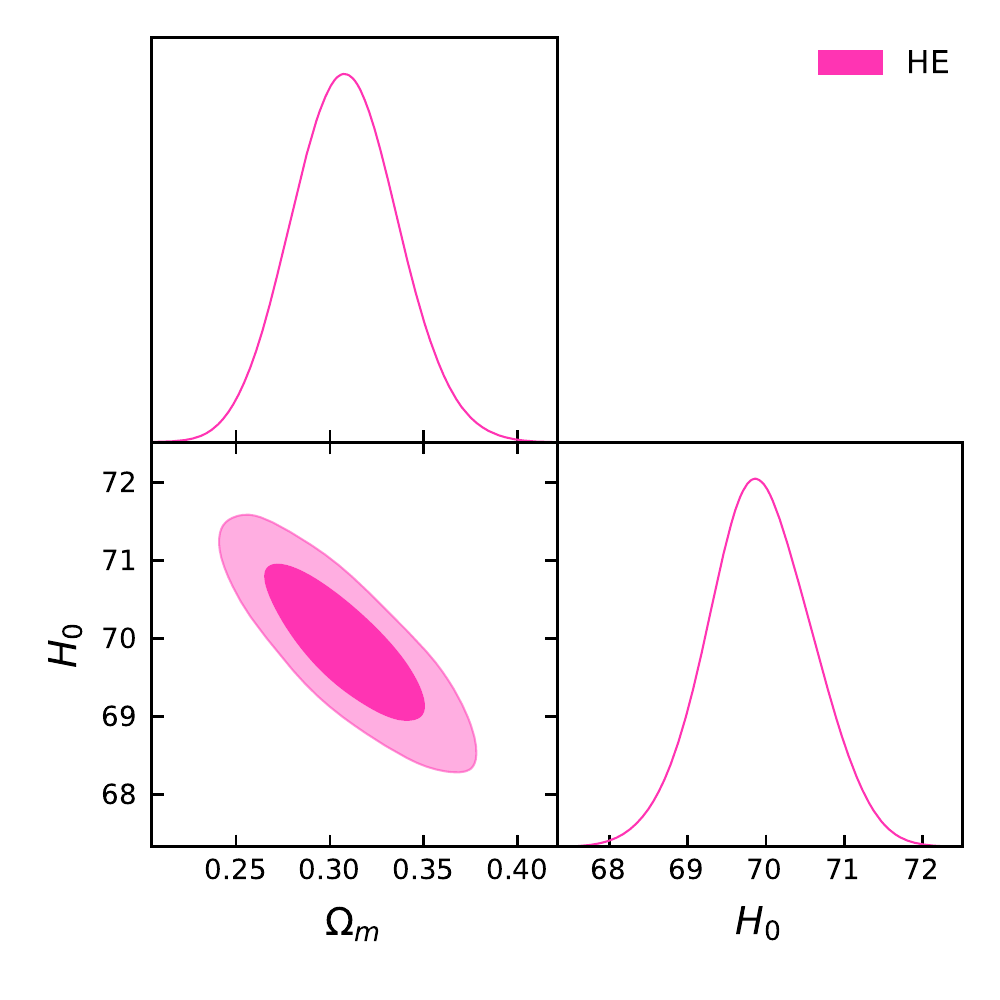}
	\includegraphics[scale=0.55]{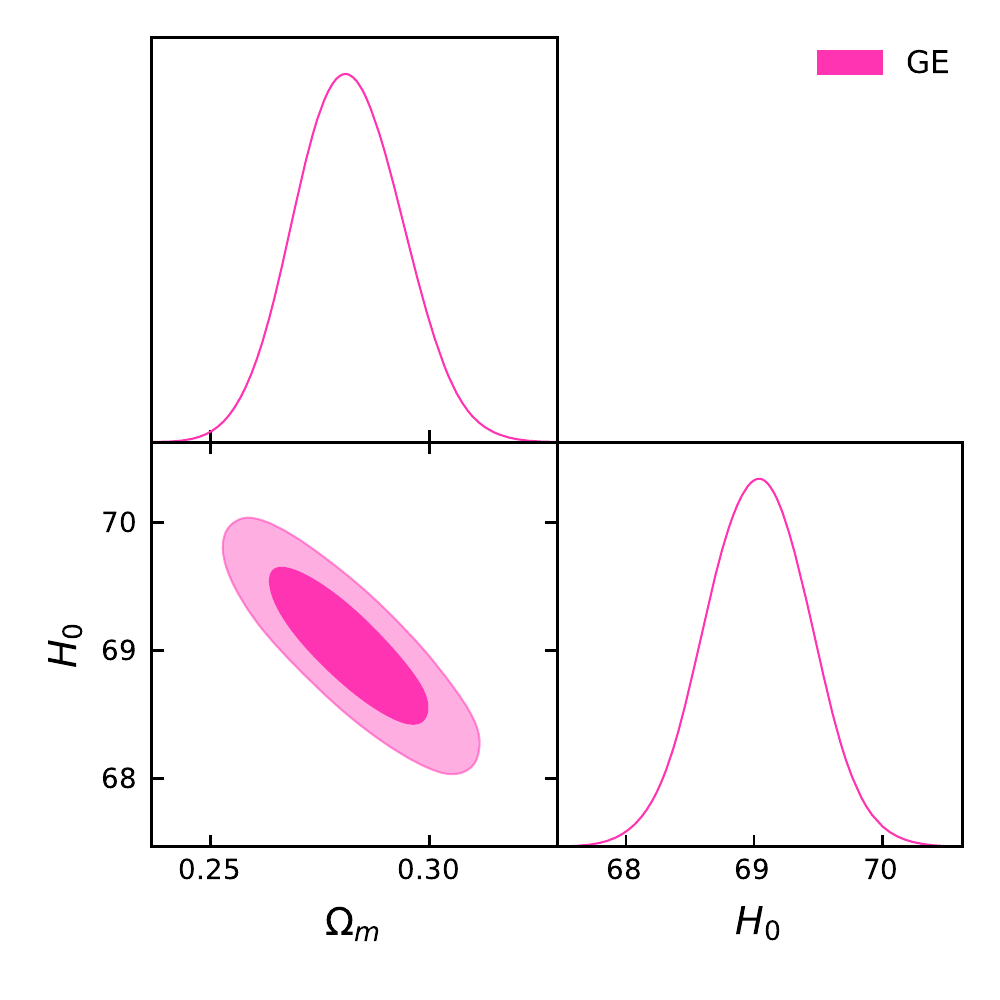}
	\caption{ The marginalized constraints on the $\Lambda$CDM model are shown by combining galaxy cluster edges data with BAO (left), CC (medium) and GW (right), respectively.}
	\label{f4}
\end{figure}

Furthermore, due to the limited constraining power of cluster edges alone, the authors in Ref.\cite{28} gives a $0.7\%$ determination on $H_0$ by combining this probe with Pantheon SNe Ia (namely SE here). It is important and interesting to study the constraining power of other probes when combined with cluster edges. 
Specifically, we constrain the $\Lambda$CDM model by combining cluster edges with CMB (CE), BAO (BE), SNe Ia (SE), CC (HE) and GW (GE), respectively. The marginalized constraining results are presented in Figs.\ref{f3}-\ref{f4} and Tab.\ref{t2}.

\begin{table}[h]
	\renewcommand\arraystretch{1.5}
	\caption{The constraining results of the $\Lambda$CDM model are shown by combining galaxy cluster edges data with CMB (CE), SNe Ia (SE), BAO (BE), CC (HE) and GW (GE) datasets, respectively.  The last two columns denote the constraining results from the combined dataset CBSHE and CBSH plus the SDSS cluster edges data, respectively.}
	{\begin{tabular}{l |c| c |c| c| c | c| c} \toprule
			Parameters            &CE           & SE      &BE          &HE           &GE                     &CBSHE        &CBSH$+$SDSS   \\ \colrule
			$H_0$              &69.07$\pm$0.36   &70.35$\pm$0.47             &$69.86\pm0.68$        &  69.92$\pm$0.67 &69.03$\pm$0.41    & 68.99$\pm$0.29   &67.98$\pm$0.39          \\  
			$\Omega_{m}$       &$0.2932\pm0.0046$  &0.289$\pm$0.012                       &$0.287\pm0.026$      &$0.319\pm0.062$  &$0.282\pm0.012$     & 0.2941$\pm$0.0038  &0.3608$\pm$0.0051                      \\
			$\Omega_{b}$       &$0.04760\pm0.00043$ &---    &---                         &---     &---                &  0.04767$\pm$0.00037      &0.04865$\pm$0.00045                       \\
			$M$                &---              &-19.343$\pm$0.013              &---    &---             &---                             &  -19.3825$\pm$0.0089  &-19.409$\pm$0.011                    \\
			\botrule
		\end{tabular}
		\label{t2}}
\end{table}

\begin{figure}[h]
	\centering
	\includegraphics[scale=0.6]{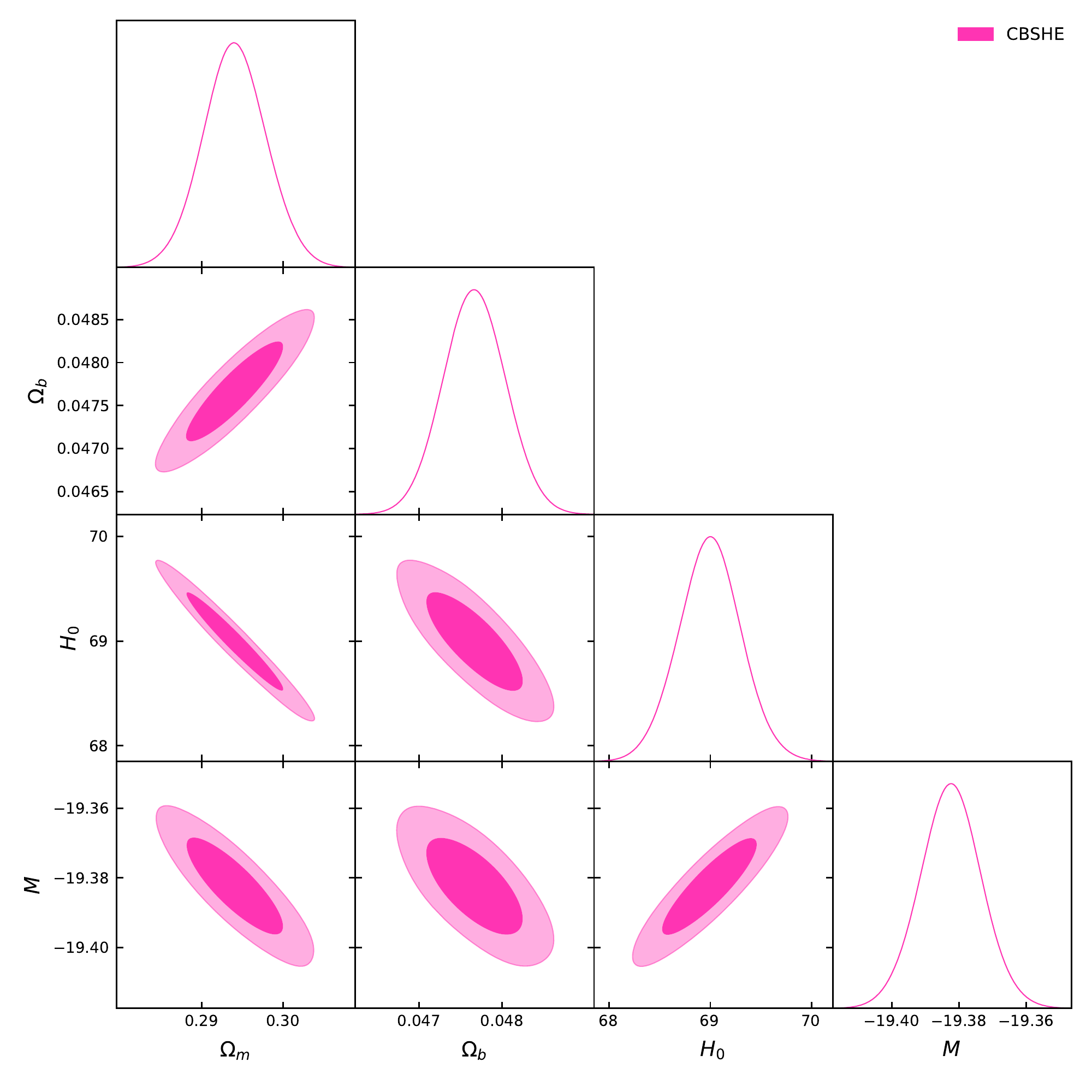}
	\caption{The marginalized constraint on the $\Lambda$CDM model is shown by using the combined dataset CBSHE. }
	\label{f5}
\end{figure}  

\begin{figure}[h]
	\centering
	\includegraphics[scale=0.6]{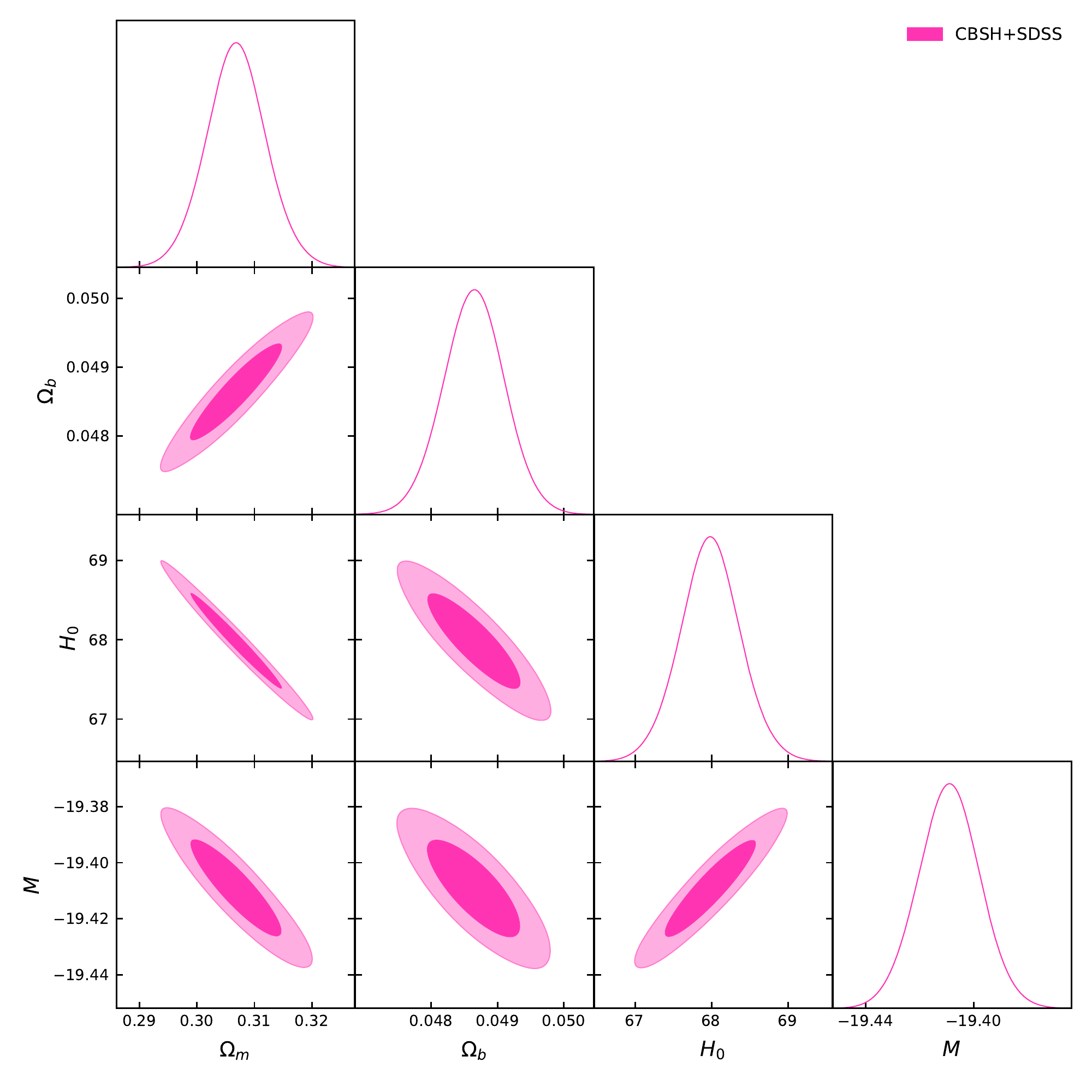}
	\caption{The marginalized constraint on the $\Lambda$CDM model is shown by combining the CBSH with SDSS cluster edges data. }
	\label{f6}
\end{figure}

For the CE case, we obtain $H_0=69.07\pm0.36$ km s$^{-1}$ Mpc$^{-1}$ with a $0.5\%$ precision. Interestingly, the accuracy of $\Omega_m$ is also improved from $2.2\%$ ($\Omega_m=0.315\pm0.007$) from the Planck-2108 CMB measurement to $1.6\%$. For the SE case, we recover the $0.7\%$ estimation of $H_0$ and gives $\Omega_m$ with an improved $4.2\%$ precision relative to $7.4\%$ estimation from the SNe Ia-only case ($\Omega_m=0.298\pm0.022$) \cite{36}. One can find the  correlations between parameters for these two combinations in Fig.\ref{f3}. It is noteworthy that cluster edges alone predicts the matter density $\Omega_m=0.325\pm0.042$ with a $12.9\%$ precision.
Employing the BAO or CC data, the precision of $H_0$ can be improved from $1.3\%$ to $1\%$. However, the BE case has a tighter estimation of $\Omega_m$ with a $9.1\%$ precision than the HE case does. Using the mocked gravitational standard sirens, we gives $H_0=69.03\pm0.41$ km s$^{-1}$ Mpc$^{-1}$, which is a $0.6\%$ estimation a little tighter than that in the SE case. Using the joint constraint from GW and cluster edges, we also obtain a $4.3\%$ estimation of $\Omega_m$. Interestingly, we find that the SE and GE cases almost share the same constraining power (see Fig.\ref{f4} and Tab.\ref{t2}).      

Finally, to estimate the parameters better, we constrain the $\Lambda$CDM model by combining current available cosmological datasets CBSH with cluster edges. The results are shown in Fig.\ref{f5} and Tab.\ref{t2}. We find that the errors of $H_0$ ($\Omega_m$) is constrained to a $0.42\%$ ($1.3\%$) precision. With the help of extra data, the constraint on the present baryon density ratio $\Omega_b$ is improved from $0.9\%$ to $0.8\%$, while the accuracy of the absolute magnitude $M$ of SNe Ia is improved from $0.07\%$ to $0.05\%$.

As a comparison, we also exhibit the tightest constraint from currently available data, i.e., CBSH plus the SDSS cluster edges data. Assuming $H_0=70$ km s$^{-1}$ Mpc$^{-1}$ , the authors in Ref.\cite{28} make full use of SDSS data and derive a $3\%$ determination on $H_0$ after calibrating the simulation (see Ref.\cite{28} for details).   
For simplicity, we add the prior $H_0=70\pm2.1$ km s$^{-1}$ Mpc$^{-1}$ into our analysis. It is worth noting that this prior is not a $H_0$ measurement from SDSS and just an estimation in light of the sensitivity of current SDSS galaxy survey. Hereafter we denote this $H_0$ prior as  ``SDSS''. 
The corresponding constraining results are shown in Fig.\ref{f6} and Tab.\ref{t2}. We find that, overall, the mocked high precision cluster edges from the DESI survey lead to a better constraint on $\Lambda$CDM than the existing SDSS data. In particular, when combined with CBSH, DESI gives $0.42\%$ determination on $H_0$, which is obviously better than $0.57\%$ precision from SDSS.

All these results imply that galaxy cluster edges from the DESI suvery can give competitive constraints on $H_0$ and $\Omega_m$ at cosmological distances relative to traditional standard probes such as SNe Ia. To constrain the cosmological models beyond $\Lambda$CDM better, cluster edges need the help of available probes.     

\section{Discussions and conclusions}
Recently, galaxy cluster edges as a standard ruler has been used to measure the cosmological distances. In this work, we go for a further step to completely assess the potential of clusters edges in constraining the cosmological models beyond $\Lambda$CDM. Specifically, we place constraints on seven cosmological models with only cluster edges data, and then constrain the $\Lambda$CDM model by combining cluster edges with CMB, BAO, SNe Ia, CC and GW, respectively.        

We find that cluster edges data from the near future DESI survey can not constrain well the typical physical parameters of the o$\Lambda$CDM, $\omega$CDM, CPL, DDE, IDE and HS $f(R)$ gravity models.  However, as we can see, the constraining precision of equation of state of dark energy from cluster edges is just about two times larger than that from the Pantheon SNe Ia sample. In addition, other probes such as CMB, SNe Ia, BAO, CC and GW, can also not give tight constraints on the typical parameters corresponding to new physics independently. Generally speaking, they are combined with each other to give better constraints. 
Therefore, we argue that cluster edges as a standard ruler can serve as a promising probe to constrain the cosmological models beyond $\Lambda$CDM in the future.     

By constraining the $\Lambda$CDM with combined datasets, we obtain the errors of $H_0$ with $0.5\%$, $0.7\%$, $1\%$, $1\%$ and $0.6\%$ precisions, uncertainties of $\Omega_m$ with $1.6\%$, $4.2\%$, $4.3\%$, $9.1\%$ and $4.3\%$ for the CE, SE, BE, HE and GE cases, respectively. Obviously, a data combination of CMB and cluster edges gives the best constraint on $H_0$ and $\Omega_m$ in these five pair datasets. Combining the gravitational standard sirens from the space-based Einstein Telescope with cluster edges, we also obtain a well constraint on $H_0$ with a $0.6\%$ precision. Very interestingly, the SE and GE pair datasets almost share the same constraining power. Using the combined datasets CBSHE, we give the most stringent constraint, which is tighter than the estimation from the CBSH plus current SDSS cluster edges data,
on the cosmological parameters in the $\Lambda$CDM model     

It is worth noting that, for cluster edges data, we have assumed fiducial $\Lambda$CDM cosmology $\Omega_m=0.3$ and $H_0=70$ km s$^{-1}$ Mpc$^{-1}$ to derive the constraints. If changing the input cosmology as Planck-2018 one $\Omega_m=0.315$ and $H_0=67.36$ km s$^{-1}$ Mpc$^{-1}$ \cite{4}, the final constraints may be slightly affected.

In order to deal with the $H_0$ tension and explore the nature of dark energy better, we hope to obtain the high precision galaxy cluster edges data by calibrating effectively the galaxy density profiles and in-fall velocity dispersion profiles. We believe that this new probe can become a complementary one relative to the traditional standard candle SNe Ia.

\section*{Acknowledgements}
Deng Wang thanks Liang Gao, Jie Wang, Qi Guo and Yun Chen for helpful discussions, and Hao-Nan Zheng, Hui-Jie Hu, Kai Zhu, Ying-Jie Jing and Hang Yang for useful communications. Deng Wang is supported by the Ministry of Science and Technology of China under Grant No. 2017YFB0203300 and National Nature Science Foundation of China under Grants No. 11988101 and No. 11851301.

\end{document}